\documentstyle[aps,twocolumn,epsfig]{revtex}
\begin{document}

\twocolumn[\hsize\textwidth\columnwidth\hsize\csname
@twocolumnfalse\endcsname

\title{Quantum-Hall Quantum-Bits}
\author{S.-R. Eric Yang$^{a,b}$, John Schliemann$^{a,c}$, 
and A.H. MacDonald$^{a}$}
\address{ $^a$Department of Physics, University of Texas, Austin, TX 78712\\
$^b$ Department of Physics, Korea University, Seoul, Korea\\
$^c$Department of Physics and Astronomy, University of Basel, Switzerland}

\date{March 2002}
\maketitle

\begin{abstract} 

Bilayer quantum Hall systems can form collective states in which
electrons exhibit spontaneous interlayer phase coherence.
We discuss the possibility of using bilayer quantum dot 
many-electron states with this property to create two-level systems
that have potential advantages as quantum bits.

\end{abstract}

\pacs{03.67.-Lx, 03.67.-a}

%% The following line should be there when using 'twocolumn'.
\vskip2pc]

Over the past several years there has been a great deal of interest in 
solid state two-level systems that could serve as quantum computing bits 
(qubits)\cite{NC},
and thereby enable large scale quantum computation.
In order for a two-level system to be useful as a quantum bit 
it must be possible to maintain coherent quantum evolution over time 
scales long 
compared to those required for its intentional manipulation.
This is particularly difficult to achieve in a solid state 
environment because of the 
presence of many degrees of freedom, electronic, photonic, and nuclear, 
that interact relatively strongly.  
The spin degree of freedom of individual electrons\cite{LoDi:98} or 
nuclei\cite{Kane} in a semiconductor, and
the total charge degree of freedom of a small superconductor\cite{Makhlin} 
are among the possibilities that have been been advanced for crafting 
solid state qubits.  In this article 
we propose a new possibility, qubits based on the the total pseudospin of
double-layer quantum dots in the quantum Hall 
regime.  We suggest an experiment which could be used to demonstrate their 
quantum coherence, and discuss some of their potential advantages. 

\begin{figure}
\center
\centerline{\epsfysize=2.3in
\epsfbox{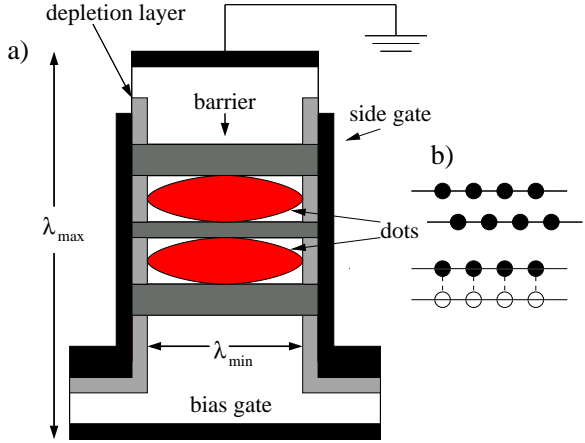}}
\begin{minipage}{8.1cm}
\caption{Schematic illustration of a pillar bilayer double-dot system. 
We consider the situation in which a strong magnetic field is applied
along the axis of the pillar.  The barriers that confine electrons 
to
the two GaAs layers in the pillar are established by the epitaxial growth 
of (Al,Ga)As layers. The maximum phonon 
wavelengths in the pillar $\lambda_{max}$ are the geometrical metrics associated with 
the structure's fundamental vibration modes, represented in this figures by the pillar height.
The minimum wavelengths $\lambda_{min}$ for phonons that couple effectively to 
collective electronic states of the dot is are comparable to the dot radius or 
the interlayer distance.  Since $\lambda_{max}$ is comparable to $\lambda_{min}$ only
the fundamental and at most of few harmonics will be important in decohering these
collective state qubits. }\label{fig:scheme}
\end{minipage}
\end{figure}

The two-level system we discuss is related to the spontaneous-coherence 
states that occur in bulk bilayer quantum Hall systems
\cite{bulkrefs,rev,wen,joglekar}.
These states have a gap for charged 
excitations that is entirely
due to electron-electron interactions.  They can be viewed 
either as pseudospin ferromagnets or as
a Bose condensate\cite{confpaper} of pairs formed between electrons in one 
layer and holes in a Landau level of an adjoining layer.  
Their low-energy physics can be described \cite{Moon} by an effective model 
usually expressed in terms of a pseudospin quantum field. The {\em pseudospin}
in these systems is the quantum {\em which-layer} degree of freedom.
An electron that is in an eigenstate of the $\hat z$-component of the 
pseudospin operator (with eigenvalue $\pm 1/2$) is definitely in one of the layers, while one 
with quantum uncertainty in its layer index is in a state with a
pseudospin component in the $\hat x-\hat y$ plane.
The qubits we propose are, however, based on {\em many-electron} 
eigenstates in small double-layer quantum dot \cite{qd} systems, like those 
realized recently by Tarucha and coworkers\cite{dd} and illustrated 
schematically in Fig.~\ref{fig:scheme}.  They have cylindrical symmetry about an axis along which a 
strong magnetic field $B\hat{z}$, can be applied.  The 
dot area is controlled by a side gate, which provides an approximately  
harmonic lateral confining potential that stabilizes finite area 
quasi-two-dimensional quantum dots.

If we assume that the quantum dots of interest are sufficiently small to suppress spatial variation 
of the pseudospin, the effective field theory \cite{Moon}  has a single 
pseudospin collective coordinate $(T_x,T_y,T_z)$ and an effective Hamiltonian:
\begin{equation}
{\cal H} = \frac{8 \pi \ell^2 \beta}{N} T_z^2 - \Delta_v T_z - \alpha \Delta_t T_x.
\label{effham}
\end{equation}
In Eq.(\ref{effham}), $T_z = -N/2,-N/2+1, \ldots, N/2$, is half the 
difference between electron numbers in right and left dots, $N$ is the total 
number of electrons in the double-dot,
$\ell$ is the magnetic length defined by $2 \pi \ell^2 B = hc/e = \Phi_0$ where
$\Phi_0$ is the magnetic flux quantum,  $\Delta_v$ is an adjustable 
interlayer bias voltage, $\Delta_t$ is the splitting of single-particle
energy levels in the double-layer system due to interlayer tunneling, $\alpha <1 $ accounts
for microscopic quantum fluctuations in the many-electron state, and 
$\beta$ is discussed below.   For $\Delta_t$ small and 
the ratio $\Delta_v N/8 \pi \ell^2 \beta$
equal to an appropriate integer, the spectrum of this Hamiltonian has two low-lying
levels separated from higher-lying states by a large gap.  This two-level system is our
proposed qubit.

\begin{figure}
\center
\centerline{\epsfysize=2.3in
\epsfbox{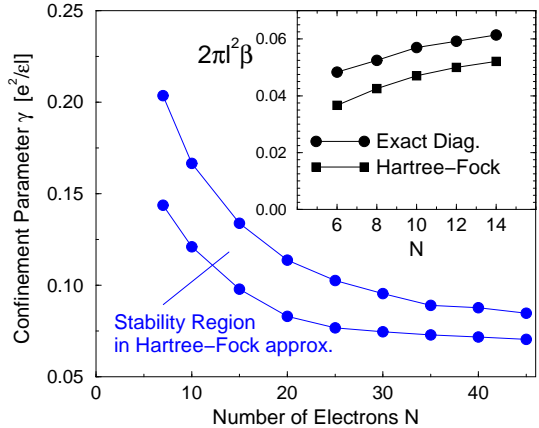}}
\begin{minipage}{8.1cm}
\caption{Hartree-Fock stability range of the bilayer coherent state
as a function of electron number $N$.  The inset displays values 
of $2\pi\ell^2\beta$ obtained by fitting the effective Hamiltonian
to exact-diagonalization and Hartree-Fock theory results.  The unit of $2\pi\ell^2\beta$
is $e^2/\epsilon\ell$.}\label{fig:one}
\end{minipage}
\end{figure}

The mean-field-theory of bilayer quantum Hall ferromagnets, like  
\cite{confpaper} the Bardeen-Cooper-Schrieffer theory of 
superconductivity,  provides both valuable insight into the nature of these 
states and a practical tool for quantitative estimates.  For bilayer 
quantum dot systems \cite{usdot,hudagotto}, the mean-field state consists of 
a Slater determinant of bilayer symmetric (for example) 
single-particle orbitals with angular momenta $m=0, \ldots, N-1$ that 
establish coherence between the layers even when 
$\Delta_t=0$.  The bilayer coherent state is stable when the occupied 
symmetric single-particle orbitals have
mean-field eigenenergies that are (i) lower than those of symmetric 
states with larger angular momenta and 
(ii) lower than those of any antisymmetric state\cite{hudagotto}.  
Note that these quasiparticle energies are split by a gap, analogous 
to the BCS gap of a superconductor, that can be due  
entirely to interactions and does not vanish for $\Delta_{t} \to 0$.  
For weak confinement or strong fields\cite{hudagotto} 
electron-electron interactions dominate and favor a fractional 
average occupation of angular momentum
states, violating the first condition.  For strong confinement or 
weak fields\cite{hudagotto} average occupation numbers larger than one are 
favored, violating the second 
condition.  As illustrated in Fig.~\ref{fig:one}  the 
double-dot coherent state is stable only over a finite range of confinement 
strengths, $\gamma=m\Omega^2\ell^2$
that narrows with increasing particle number $N$ ($\Omega $ 
is the frequency of the harmonic potential). 
The value of $\beta$ can be estimated, using either mean-field or exact 
diagonalization calculations, from the dependence of the ground state pseudospin polarization 
on bias potential.
From the inset of Fig.~\ref{fig:one} we see that $2 \pi \ell^2 \beta = 0.06 e^2/\epsilon \ell$
at $N=14$, compared to the bulk value $2 \pi \ell^2 \beta = 0.09 e^2/\epsilon \ell$ \cite{Moon}.
The results of Fig.~\ref{fig:one} were obtained with the interlayer separation equal to $\ell$ and 
with $\Delta_t$ and  $\Delta_v$ set to zero.  

\begin{figure}
\center
\centerline{\epsfysize=2.3in
\epsfbox{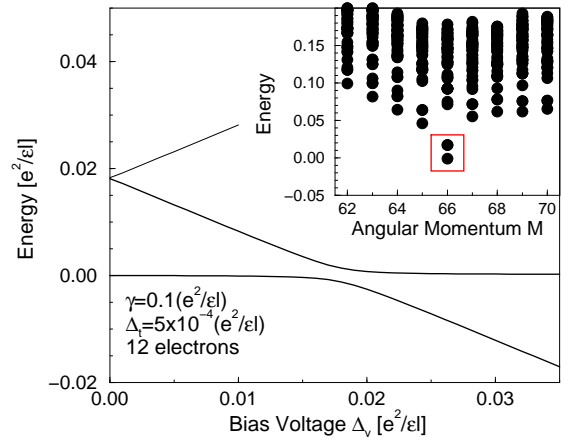}}
\begin{minipage}{8.1cm}
\caption{Dependence of the three lowest eigenvalues with total 
angular momentum $M=M_{GS}$ on bias voltage
$\Delta_v$. The inset shows the low-energy part of 
the many-electron energy spectrum [in units of
$e^2/\epsilon\ell$] as a function of 
total angular momentum $M$ for 12 electrons at $\Delta_v=0$ and at the 
strength of the confinement potential $\gamma=0.1e^2/\epsilon\ell$.
$M_{GS}=66$ for $N=12$.  This figure demonstrates that the two-level system
of our proposed qubit is cleanly separated from other many-electron states.
}\label{fig:two}
\end{minipage}
\end{figure}

In the inset of Fig.~\ref{fig:two} we plot the exact spectrum of the 
many-particle Hamiltonian for 12 electrons in a double-layer quantum dot system as a 
function of total angular momentum $M$. 
The collective state discussed above has quasiparticles with angular momentum 
$m = 0, \ldots, N-1$ and so occurs at total angular momentum 
$M_{DLCS}=N (N-1)/2$.  Exact diagonalization calculations show that the double-layer 
coherent state is the ground state over a narrow range of field 
strengths. The results plotted in Fig.~\ref{fig:two} are for $\gamma=0.1e^2/\epsilon\ell$, which is 
in the middle of the stability range for $N=12$.  Comparing with Fig.~\ref{fig:one}, we see that 
Hartree-Fock theory only slightly overestimates the confinement strength at which the double-layer
coherent state occurs.  The main part of Fig.~\ref{fig:two} shows the 
dependence of the three lowest eigenvalues on $\Delta_v$ for 
$\Delta_t$ equal to a small positive value.  
The first anticrossing takes place near $\Delta_v=0.02e^2/\epsilon\ell$, 
consistent with the effective model prediction, 
$\Delta_v = 8 \pi \ell^2 \beta/N$.
These results demonstrate that the effective Hamiltonian accurately describes
the low-energy part of the dependence of the microscopic Hamiltonian's 
spectrum on $\Delta_t$ and $\Delta_v$, which will figure importantly in our discussion.
The lowest energy states that appear at $M > M_{GS}$ in Fig.~\ref{fig:two}
are edge excitations of the dot, which form a part of our proposed qubit's 
environment as discussed below.

For $N$ even, assumed in the following, we choose 
the many body states $|0\rangle$ and  $|1\rangle$  ($T_z$  
eigenvalues are 0 and $1$) as the orthogonal states of the proposed qubit.
(For $N$ odd, the optimal orthogonal states would be $|1/2\rangle$ and  $|-1/2\rangle$.)  
At zero bias voltage the groundstate is non degenerate and, for small 
$\Delta_t$ has nearly definite $T_z$.  
The truncated two-level Hamiltonian is given up to a constant by 
\begin{equation}
H_P=\left(\frac{4\pi\ell^2\beta}{N}-\frac{\Delta_v}{2}\right) 
\sigma_z- \frac{\Delta}{2}\sigma_x,
\label{eff}
\end{equation}
where $\Delta = \alpha \Delta_t \sqrt{(N+2)N} /2$, 
and $\sigma_z$ and $\sigma_x$ are Pauli spin matrices in the space of the qubit.
The tunneling term in the Hamiltonian, which is very small for the quantum dot
systems we have in mind, is effective only near the ${\em resonance}$ situation,
$\Delta_v = 8 \pi \ell^2 \beta/N$, where it splits the degeneracy which would otherwise occur.
From Fig.~\ref{fig:two} we find that 
$\Delta=2.57\times 10^{-3}e^2/\epsilon\ell$, corresponding to  
$\alpha=0.86$. 

In this paragraph we propose an experiment which we believe is feasible and 
which can be used to study the coherent quantum evolution of this qubit. 
If the bias voltage is held at zero, where the $\hat z$ component of the qubit effective 
field dominates, for a sufficiently long time the qubit will reach its ground state, $|0\rangle$.
When the bias voltage is tuned to resonance, 
the qubit evolves under the influence of the $\hat x$ direction Zeeman field. 
If after a time $T$ the bias voltage returns to zero, the qubit will have a 
finite probability amplitude for being in the $|1\rangle$ state, {\it i.e.} 
for having transferred an electron between layers of quantum dot.
We require that equilibrium be reestablished by charge 
flowing through the external circuit used to control the bias voltage, 
a property that requires the dot contact resistances to be smaller than the 
off-resonance intra-dot resistance. 
If this procedure is repeated many times with a repetition period 
$T_{repeat}$ that is longer than the equilibration time, the current flowing 
between layers through the external circuit measures the probability for transferring an electron:
\begin{equation}
I = e\left(\frac{1}{2}-\frac{1-2p}{2}\cos(\Delta T/2\hbar)\right)/T_{repeat}.
\label{eq:iavg}
\end{equation}
Here $p \ne 0$ is small if the resonance condition is established rapidly compared to 
$\hbar /\Delta$ as we discuss below.
A measurement of oscillating current as a function of time-on-resonance 
$T$ would establish quantum coherence for this potential qubit.

There are, of course, no genuine two-level systems in nature, 
and the one under discussion
here is not an exception.  Even discounting coupling to 
phonons, nuclear spins, and other 
degrees of freedom present in the solid state environment, 
the {\em electron} Hamiltonian 
has a large number of eigenstates.  A practical requirement\cite{adia} 
for an effective two-level system is that  
its operational time scale , $\hbar /\Delta$,
be long compared to the scale which breaks adiabaticity 
between the two-level system
and its electronic environment.  If the pseudospin effective 
Hamiltonian fully described
the double layer quantum-dot, this time scale would be $\hbar / 
(2 \pi \ell^2 \beta)$.
If rotational symmetry of the Hamiltonian is broken, on the other hand,
edge states are the lowest lying electronic 
excitations outside of the two-level system.  
As illustrated in Fig.~\ref{fig:two}, the size of these energies for 
realistic double-layer quantum dots is $\sim 1.0$meV for 
$e^2/\epsilon\ell \sim 10$meV.  
The tunneling amplitude between 
bilayer systems can be varied over a very wide range because of its 
exponential dependence
on Al content in the barrier separating the quantum wells and on the 
width of this barrier.  
% Question: I think that we need to discuss this. 
For $\Delta_{t}=0.0003$meV and $N=40$ the gap $\Delta$ is  
0.001 meV, safely smaller than edge excitation energies, and 
the operational time of the quantum bit is $\sim 10^{-9}$ s, 
making it possible to adjust the bias potential sufficiently rapidly
to achieve $p\ll 1$ in Eq.(~\ref{eq:iavg}).

\begin{figure}
\center
\centerline{\epsfysize=2.3in
\epsfbox{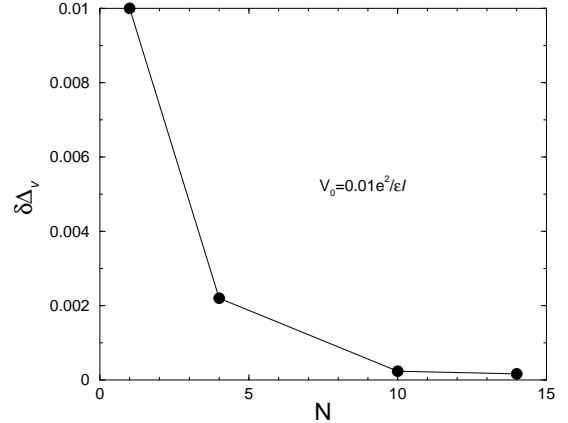}}
\begin{minipage}{8.1cm}
\caption{Dependence of the change in the qubit parameter $\Delta_v$ on $N$ in the presence of
a fluctuation in the potential
difference between the two layers that is localized at the center of the quantum dot.
This quantity has been evaluated by performing self-consistent Hartree-Fock calculations
at $\Delta_t=0$ and evaluating the energy difference between states that differ by one in 
the number of electrons in the top layer.  The potential fluctuation strength has been 
chosen so that it alters $\Delta_{v}$ by 0.01 $e^2/\epsilon \ell$ for a single-electron.
Electronic correlations in large $N$ collective states, suppress the system's response to
the fluctuation potential, and reduces the change in $\Delta_{v}$ that results from the same
potential fluctuation.
}\label{fig:four}
\end{minipage}
\end{figure}

Bilayer quantum Hall systems are similar\cite{wen} in many respects to 
Josephson coupled superconductors (but also show differences
\cite{joglekar}), and there are both similarities and 
differences between
the experiment we propose above and the coherent single-Cooper-pair box 
effect realized\cite{nakamura}
by Nakamura {\it et al.}.  In the Nakamura experiment, the qubit is formed by 
states that differ by 1 in the number of Cooper pairs on one side of the 
junction, whereas in
the excitonic insulator language our qubit is formed by states that 
differ by one in the 
number of electron-hole pairs in the system.  Since a new 
electron-hole pair is formed by
moving a single electron from one-layer to the other, the current 
expression in our case is smaller by a factor of two.  The two qubits share 
the property that, 
because of the collective behavior of many electrons, coherence can be 
established even though the systems are composed of a relatively large number 
of electrons.  
The phase coherence time of bulk bilayer quantum Hall systems has been 
estimated\cite{rev} to be $\tau_{\phi}\sim 2\times 10^{-10}$ s,
most likely\cite{joglekar} due to coupling between collective degrees of 
freedom and electronic excitations near the boundaries between the 
incompressible and compressible regions that occur in all bulk samples.
This coherence time is shorter than the period we propose above for a typical
Rabi-oscillating current made from quantum Hall quantum bits.  
However, as explained above, this decoherence mechanism
is not operative for quantum dot systems,  Instead the dominant decoherence
mechanism is likely to be piezoelectric coupling to long-wave length phonons, 
nabbed\cite{Fuji} as the likely culprit for single-electron double-dots\cite{Fuji,bayer}.  
For collective qubits, like the ones we propose, only phonons with wavelengths 
comparable to the entire electronic system can effectively decohere. 
We illustrate this point in Fig.~\ref{fig:four} by plotting the dependence 
of the change in $\Delta_{v}$  on $N$ in the presence of a 
potential fluctuation located at the center of the quantum dot.
% Eric: I think that we should write the vertical axis as $\delta \Delta_{v} and the 
% and divide all the numbers by 2?   
For large $N$, correlations between 
electrons in the collective state suppress response response to the localized
potential fluctuation, and  limit its effectiveness in decohering the qubit.
We expect that phonons will be an especially weak decoherence mechanism
for these qubits because the maximum phonon wavelength, established by size 
quantization, is only slightly larger than the minimum effective phonon 
wavelength, established by the size of the double-dot system.  The fact that the
decoherence time is already $\sim 10^{-10}$ s in the bulk, suggests that very long coherence times will be 
achievable.

Solid state quantum computing clearly presents 
even more daunting obstacles than merely coherent time evolution of solid 
state qubits.  The problems which would need to be solved to 
control the entanglement of these quantum Hall qubits 
are similar to entanglement strategies for the case of of Cooper pair
box qubits\cite{Makhlin}.  We believe that experimental study of the novel
qubit proposed here, which occurs in a physical system that has
already been achieved\cite{dd}, could make 
valuable contribution to the growing understanding of
decoherence in solids.  

We are grateful for valuable 
conversations with G. Austing, S. Tarucha, D. Loss, Y. Nakamura, and P. Zoller.
S.R.E.Y was supported by the KOSEF Quantum-functional Semiconductor 
Research Center at Dongguk University, J.S. was supported by the 
Deutsche Forschungsgemeinschaft.  Work at the University of Texas was supported
by the Welch Foundation and the by the National Science Foundation under grant DMR0115947.

\end{document}